\def\tSW{\left \langle \tau_{\textsc{sw}} \right \rangle}
\def\dKC{d_{\textsc{kc}}}
\def\tdKC{\widetilde{d}_{\textsc{kc}}}
\def\gKr{\gamma_{\text{Kr}}}

\documentclass[aps,prappl,twocolumn,groupedaddress,showkeys,showpacs,superscriptaddress,floatfix]{revtex4-2}
\usepackage{listings}
\usepackage{epsfig}
\usepackage{multirow}
\usepackage{amsmath, amssymb,mathtools}
\usepackage{xcolor}
\usepackage{graphicx}
\usepackage{dcolumn}
\usepackage{bm}
\usepackage{nicefrac} 
\usepackage{subfigure}
\usepackage[utf8]{inputenc}
\usepackage[T1]{fontenc}
\usepackage{tikz}
\usepackage{orcidlink}

\begin{document}

\title{Josephson-based scheme for the detection of microwave photons}

\author{Claudio Guarcello\,\orcidlink{0000-0002-3683-2509}}
\email{Corresponding author: cguarcello@unisa.it}
\affiliation{Dipartimento di Fisica ``E.R. Caianiello'', Universit\`a di Salerno, Via Giovanni Paolo II, 132, I-84084 Fisciano (SA), Italy}
\affiliation{INFN Gruppo Collegato di Salerno, Via Giovanni Paolo II, 132, I-84084 Fisciano (SA), Italy}
\author{Alex Stephane Piedjou Komnang\,\orcidlink{0000-0003-2430-0086}}
\email{apiedjoukomnang@unisa.it}
\affiliation{Dipartimento di Fisica ``E.R. Caianiello'', Universit\`a di Salerno, Via Giovanni Paolo II, 132, I-84084 Fisciano (SA), Italy}
\author{Carlo Barone\,\orcidlink{0000-0002-6556-7556}}
\email{cbarone@unisa.it}
\affiliation{Dipartimento di Fisica ``E.R. Caianiello'', Universit\`a di Salerno, Via Giovanni Paolo II, 132, I-84084 Fisciano (SA), Italy}
\affiliation{INFN Gruppo Collegato di Salerno, Via Giovanni Paolo II, 132, I-84084 Fisciano (SA), Italy}
\affiliation{CNR SPIN Salerno, Via Giovanni Paolo II, I-82084 Fisciano (SA), Italy}
\author{Alessio Rettaroli\,\orcidlink{0000-0001-6080-8843}}
\email{alessio.rettaroli@lnf.infn.it}
\affiliation{INFN, Laboratori Nazionali di Frascati, Frascati (Roma) Italy}
\affiliation{Dept. of Mathematics and Physics, University of Roma Tre, I-00100 Roma, Italy}
\author{Claudio Gatti\,\orcidlink{0000-0003-3676-1787}}
\email{claudio.gatti@lnf.infn.it}
\affiliation{INFN, Laboratori Nazionali di Frascati, Frascati (Roma) Italy}
\author{Sergio Pagano\,\orcidlink{0000-0001-6894-791X}}
\email{spagano@unisa.it}
\affiliation{Dipartimento di Fisica ``E.R. Caianiello'', Universit\`a di Salerno, Via Giovanni Paolo II, 132, I-84084 Fisciano (SA), Italy}
\affiliation{INFN Gruppo Collegato di Salerno, Via Giovanni Paolo II, 132, I-84084 Fisciano (SA), Italy}
\affiliation{CNR SPIN Salerno, Via Giovanni Paolo II, I-82084 Fisciano (SA), Italy}
\author{Giovanni Filatrella\,\orcidlink{0000-0003-3546-8618}}
\email{filatrella@unisannio.it}
\affiliation{Dep. of Sciences and Technologies, University of Sannio, Via De Sanctis, Benevento I-82100, Italy}
\affiliation{INFN Gruppo Collegato di Salerno, Via Giovanni Paolo II, 132, I-84084 Fisciano (SA), Italy}
\date{\today}

\begin{abstract}
We propose a scheme for the detection of microwave induced photons through current-biased Josephson junction, from the point of view of the statistical decision theory.
Our analysis is based on the numerical study of the zero voltage lifetime distribution in response to a periodic train of pulses, that mimics the absorption of photons.
The statistical properties of the detection are retrieved comparing the thermally induced transitions with the distribution of the switchings to the finite voltage state due to the joint action of thermal noise and of the incident pulses.
The capability to discriminate the photon arrival can be quantified through the Kumar-Caroll index, which is a good indicator of the Signal-to-Noise-Ratio.
The index can be exploited to identify the system parameters best suited for the detection of weak microwave photons.
\end{abstract}

\maketitle

\section{Introduction}
\label{Sec00}\vskip-0.2cm
The detection of rare and weak electromagnetic signals finds application in many fields, e.g. gigahertz astronomy and axions search~\cite{asztalos2010,Beck13,graham2015}, quantum computing~\cite{OBr07} and superconducting electronics~\cite{Gu17,Bra19,Abd20}, for which very fast, low noise, and extremely sensitive detectors in the microwave-to-Terahertz range are required.
Indeed, in recent years the progresses of nanotechnologies for superconducting devices have led to the design of highly sensitive superconducting detectors for electromagnetic radiation, such as transition edge sensors~\cite{AlesiniBabusci20,Alesini20}, kinetic inductance detectors~\cite{Zmu12}, Josephson escape sensors~\cite{PaoBuc20} and travelling-wave parametric
amplifiers~\cite{Mac15,Gri21}, graphene-\cite{Walsh2017,Kok20,Lee20,Wal21} and qubit-based detectors~\cite{Bes18,Kon18,Das20},  high-sensitivity calorimeters~\cite{Guarcello2019_1,Guarcello2019_3,PaoGiaz21} and cold-electron bolometers~\cite{AngKuz20}. The latter have proved suitable for detecting low-energy, e.g., 30 GHz, photons generated by rare events, such as axion conversion, due to a fast and noise-protected response~\cite{Gia06,AngKuz20}.

In this scenario, Josephson junctions (JJs) are promising single-photon, current-controlled, threshold sensors that operate at cryogenic temperatures~\cite{Yab21,Gol21}. Moreover, JJs can be easily integrated in large numbers, thus allowing the scaling to array detection systems. 

The JJ detection process can be investigated through the study of  the dynamics of a ``virtual particle'', whose position is given by the Josephson phase, subject to a cosine-``washboard'' potential in which it is confined~\cite{Barone82}. The zero voltage state of a JJ corresponds to such particle trapped in one well of the washboard potential. 
The escape of the particle out of the potential well towards the running state, after which the particle slides down along the potential slope, corresponds to the appearance of a measurable voltage across the JJ. 
Such escape can be determined by the absorption of an electromagnetic signal~\cite{Filatrella10,Chen11,Oel13,Oel17,Kuzmin18,Gua20,Anghel2020,Revin20}, or by thermal activation above the well energy barrier due to intrinsic noise~\cite{Kramers40}, or even by macroscopic quantum tunneling~\cite{Dev85,Cla88}.
By repetitively measuring the lifetime of the zero voltage state it is possible to retrieve its statistical distribution, and from a further analysis it is possible to reveal the presence of an exciting signal, even in the presence of intrinsic noise~\cite{Pie21}.

In this work we numerically investigate, in the framework of signal detection~\cite{Filatrella10,Addesso13-2,Addesso16}, a scheme to reveal the absorption of photons through the analysis of the switching time distributions in a JJ~\cite{Pie21,Fil21}.
In particular, we discuss how to plan Josephson detection experiments to reveal the existence of a weak signals in a thermal noise background. The employed methodology consists in comparing the switching probabilities of a current biased JJ exposed to a train of small current pulses, which mimics a weak photon field, with that obtained in the absence of signal. The investigation of the unbalance in the distribution of switching times in the two cases gives an estimate of the efficiency of the detection.
The performances of the detector are quantitatively evaluated through the Kumar-Caroll (KC) index~\cite{Kumar84,Filatrella10,Addesso12,Pou20}, which is a proxy for the Signal-to-Noise-Ratio (SNR). The optimization of the detection probability can provide a guide to select the JJ parameters that best suit the detection of weak microwave signals.

Many noisy systems, whose response is characterized by a sudden passage from an initial metastable state to a readable outcome, have been already studied through the KC index, \emph{e.g.}, from Josephson devices~\cite{Filatrella10,Addesso12,Pou20} to Fabry-Perot interferometers~\cite{Addesso13-1,Addesso15}.
In these cases, one usually measures the escape times, or even the switching currents dealing with Josephson devices, to retrieve some information about the physical system. The combined action of the bias current and noisy fluctuations has already demonstrated a peculiar response in the distribution of Josephson switching times~\cite{Gua15} and currents~\cite{Guarcello2017,Guarcello19}, both in the case of Gaussian and non-Gaussian noise sources~\cite{Gua13,Gua14,Guarcello16}.

The paper is organized as follows. Section~\ref{Sec01} is devoted to the description of the JJ model used. Section~\ref{Sec01-A} is dedicated to introduce the details of the detection strategy and the KC index. In Section~\ref{Sec03} the numerical computational details are presented. In Section~\ref{Sec02} we discuss the obtained results and the detection performances as a function of the JJ parameters. Finally, in Section~\ref{Sec04} the conclusions are drawn.

\section{Model}
\label{Sec01}\vskip-0.2cm

\begin{figure}[t!!]
\includegraphics[width=\columnwidth]{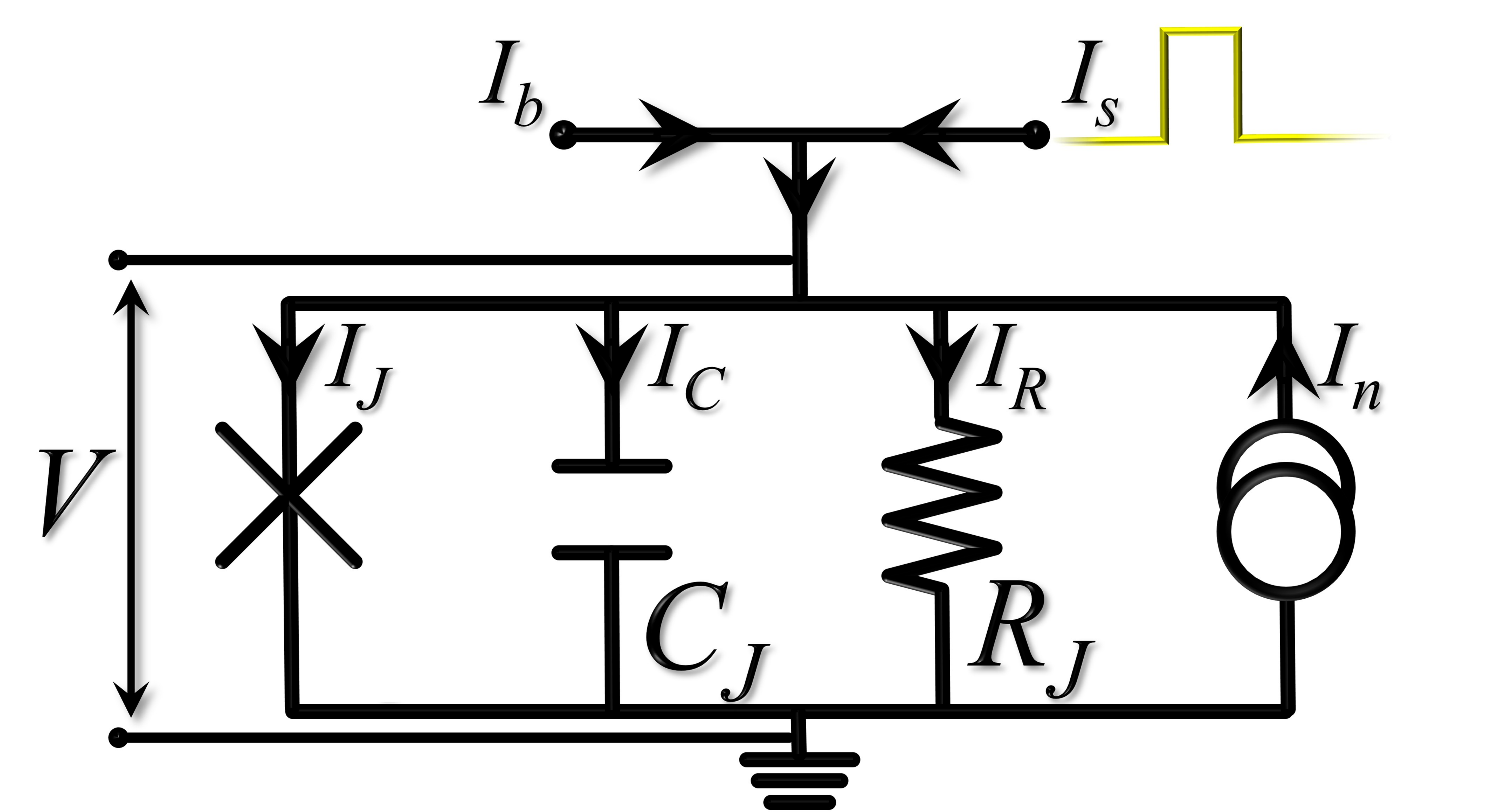}
\caption{Equivalent electrical model of a current-biased short Josephson junction. The dc bias current, $I_b$, the current source used to mimic a weak microwave field, $I_s$, and the noise current, $I_n(t)$, of the JJ are included in the diagram. $R_J$ represents the quasiparticle tunneling resistance and $C_J$ the junction capacitance.}
\label{Fig01}
\end{figure}

The electrodynamics of a short tunnel type JJ can be described in terms of the Resistively and Capacitively Shunted Junction (RCSJ) model~\cite{Barone82}. This is an electrical model whose elements are related to specific physical characteristics of the junction, \emph{i.e.}, the capacitance $C_J$ between the JJ electrodes of the device, the resistance $R_J$, due to quasiparticles tunneling, and $I_J$ the Josephson supercurrent. The electrical currents flowing through the junction are provided by external (the constant bias current and the electromagnetic signal to be detected) and intrinsic (the Johnson noise associated to quasiparticle tunneling) sources, see Fig.~\ref{Fig01}.

The Josephson current $I_J$ and the corresponding voltage drop $V$ depend on the phase difference, $\varphi$, between the macroscopic wavefunctions of the two superconductors, following to the Josephson equations~\cite{Jos62,Jos74}
\begin{equation}\label{JJvoltage}
I_J = I_0\sin\varphi \qquad\text{and}\qquad V = \frac{\hbar}{2e}\frac{d\varphi}{dt}.
\end{equation}
Here, $I_0$ is the critical current, that is the maximum supercurrent that can flow in the JJ element, and $\hbar$ and $e$ are the reduced Planck constant and elementary charge, respectively.
If a dc bias current, with a value below $I_0$, is present, the Josephson equations admit as a solution a stationary phase,  corresponding to a zero average voltage. Conversely, if the dc bias current overcomes $I_0$, Eq.~\eqref{JJvoltage} predicts an ever increasing phase, that corresponds to a finite-voltage state.
The resistor $R_J$  depends nonlinearly both on voltage and temperature, although, as it is related to the dissipation in the system,  in the case of a moderate or weak damping its nonlinearity is often not taken into account.
Associated to the Ohmic resistor $R_J$ there is a thermal noise current source, indicated with $I_n$ in Fig.~\ref{Fig01}, whose spectral power density is frequency-independent but temperature-dependent (Johnson noise)~\cite{Kogan96,Barone18,Barone21}.

\begin{figure}[t!!]
\includegraphics[width=\columnwidth]{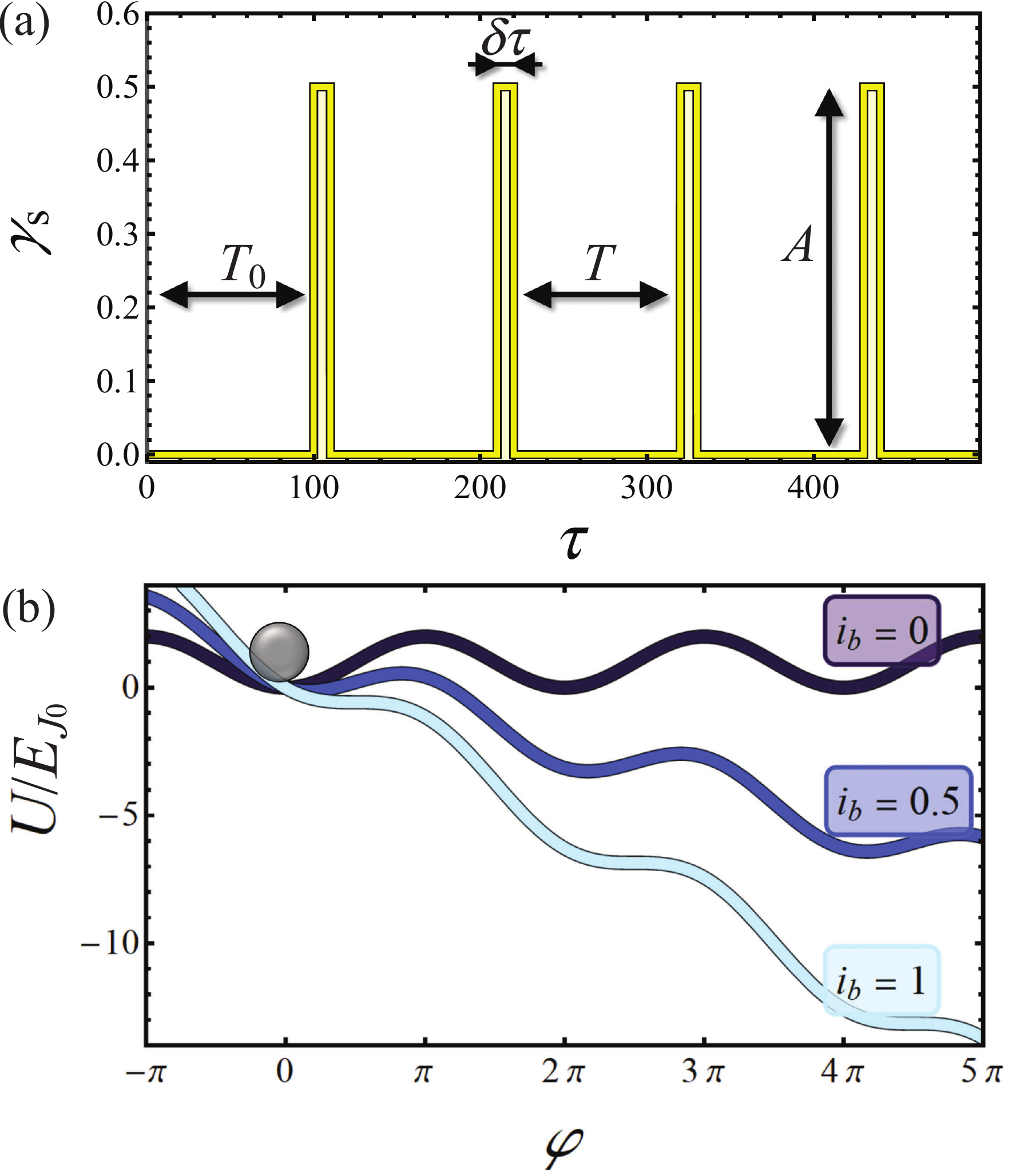}
\caption{(a) Train of current pulses used to mimic the effect of photons absorption. The height and the width of each pulse, $A$ and $\delta t$, the distance, $T$, between two consecutive pulses, and the time $T_0$ the first pulse takes to arrive are also indicated. Each experiment consists in $N$ trains of pulses, each sequence spans for a time $P$. (b) Sketch of the tilted washboard potential in the case of three different bias current values.}
\label{Fig02}
\end{figure}

The external signal to be detected is a weak microwave field coupled to the junction (``weak'' because carrying few photons). It is modeled as a deterministic source, labeled with $I_s$ in Fig.~\ref{Fig01}, of well separated, periodic, rectangular current pulses, each  corresponding to an absorbed photon, see Fig.~\ref{Fig02}(a).

The detection experiment considered in this work involves a JJ initially d.c. biased in the zero voltage state by a current $I_b$ smaller than the critical value $I_0$. Subsequently, an external signal, generated by $I_s$, is applied and the appearance of a finite voltage is monitored to determine the value of the zero voltage state lifetime. In the underdamped case, it is unlikely that once it has escaped, the particle will be trapped again: the phase reaches a constant speed, corresponding to a voltage step, until the superconducting regime is established again by a change in the current bias.
Moreover, the presence of thermal noise $I_n$ alone can cause a premature switching to the finite voltage state, that, in the detector jargon, is called \emph{dark count}. Of course, the distinction between real and dark counts cannot occur on the basis of a single event, therefore a statistical analysis of the switchings distribution is necessary.
Let us add, for sake of completeness, that the switchings can be collected in a variety of detection schemes, \textit{e.g.}, biasing the junction with an oscillating~\cite{Addesso12,Gua15,Yab20} or a linearly ramping electric current~\cite{Filatrella10,Pou20}.

Thus, considering all the current contributions and Josephson equations, from Kirchhoff's laws it is straightforward to derive the following Langevin equation~\cite{Ben84}
\begin{equation}\label{eqJJ}
C_J\frac{\hbar}{2e}\frac{d^2\varphi}{dt^2}+\frac{1}{R_J}\frac{\hbar}{2e}\frac{d\varphi}{dt}+I_0\sin\varphi = I_b + I_s\left(t \right) + I_n\left(t \right).
\end{equation}
In terms of the normalized time $\tau=\omega_J t$, with $\omega_J=\sqrt{2eI_0 / C_J\hbar}$ being the Josephson plasma frequency~\cite{Barone82}, Eq.~\eqref{eqJJ} can be cast as
\begin{eqnarray}\label{eqJJ_norm}
\frac{d^2\varphi}{d\tau^2}+\alpha\frac{d\varphi}{d\tau}+\sin\varphi = \gamma_b + \gamma_s\left(\tau \right) + \gamma_n ({\tau}).
\end{eqnarray}
In this, we recognize the damping parameter $\alpha = 1/(R_J C_J\omega_J)$, the normalized currents $\gamma_b=I_b/I_0$ and $\gamma_s = I_s/I_0$, and the stochastic noise term $\gamma_n=I_n/I_0$, which has the following statistical properties
\begin{equation}\label{correl}
\langle \gamma_n(\tau) \rangle = 0 \quad\text{and} \quad
\langle \gamma_n(\tau),\gamma_n(\tau') \rangle = 4 \alpha\, D\, \delta(\tau-\tau').
\end{equation}
Here, the parenthesis $\langle \rangle$ indicate ensemble averages, $\delta()$ is the Dirac delta function, and
\begin{equation}\label{noiseampl}
D=\frac{2e}{\hbar}\frac{k_B {\cal T}}{I_0}
\end{equation}
is the normalized noise intensity, where $k_B$ is the Boltzmann constant and $\cal{T}$ is the JJ temperature. 
The normalized values used in our calculations for the damping parameter and the noise amplitude are $\alpha \simeq 0.025$ and $D\simeq0.01$, respectively. 
These values correspond to an Al-based JJ having a normal-state resistance $R_J\simeq0.57\;\text{k}\Omega$, a capacitance $C_J\simeq3.2\;\text{pF}$, a critical current $I_0 \simeq 0.5\;\mu \textup{A}$, a Josephson frequency $\omega_J \simeq 22\;\textup{GHz}$, a critical temperature $T_c=1.2\;\text{K}$, and operating at ${\cal{T}} \simeq 0.12\;\text{K}$.

In Fig~\ref{Fig02}(a) is shown a train of current pulses that mimics the arrival of photons and that can be expressed as
\begin{equation}\label{gamma_ac}
\gamma_s(\tau) = \sum_{n=-\infty}^{+\infty}A \Big[ \theta \big( \tau - nT \big) - \theta \big( \tau - n(T+\delta \tau) \big) \Big],
\end{equation}
where $\theta$ is the Heaviside step function, $A$ is the normalized amplitude of the signal, $\delta \tau$ is the pulse width, and $T$ is the distance between consecutive pulses arriving with the repetition rate $r_A = 1/T$.
The initial time $T_0$ prior to the arrival of the very first pulse to the junction is a random time uniformly distributed between $0$ and $T$. We observe that when the time interval between two ``photon packets'' is large enough (that is, $T\gg 1/\alpha = 50$ in our case), the packets do not interfere with each other; moreover, a random distribution of the distance between two packets whose dispersion is much less than the average distance would not alter the results we present in this work.
Finally a negative value ($A<0$) of the amplitude of the wave-packet would depress the average escape time, as the energy barrier would increase. However, because of the symmetry of the problem, the same effects discussed in the following could be obtained reversing the direction of the bias current in Eq.\eqref{eqJJ}.

To give a realistic measure of the photon signal amplitude, we observe that in Ref.~\cite{Oel17} the estimated mean amplitude of the current produced by a photon absorption is $I_{ph}\approx25\;\text{nA}$. This quantity depends on the capacitance and inductance of the resonator, and also on the inductance of the connected JJ, which in turn is a function of both the bias and the critical currents. In our normalized units and assuming the same resonator's parameters used in Ref.~\cite{Oel17}, the normalized photon current amplitude reads $A_{ph}\approx 0.05$~\footnote{For the sake of completeness, we observe that in Ref.~\cite{Oel17} the photon is assumed to induce an rf current, while in our work we consider rectangular current pulses. The effects connected to the specific shape of the pulses will be the subject of future works.}. However, for the sake of completeness, we have considered amplitudes in a quite large range of values, \emph{i.e.}, $A\in[0.1-0.5]$. This allows  to discuss the system response in both the low- and high-amplitude regimes. In the former, the pure-noise and the pulse-sustained switching dynamics are quite close, just like in a real single-photon detection experiment, while in the latter the pulse-sustained evolution departs significantly from the pure-noise situation.

The washboard potential responsible for the phase dynamics $U(\varphi)$, see Fig.~\ref{Fig02}(b), has the functional form~\cite{Barone82, Ben84}
\begin {equation}\label{eq:potential}
U(\varphi,\gamma) = E_{J_0}[1 - \cos (\varphi) - \gamma\; \varphi],
\end {equation}
where $E_{J_0}=\left (\hbar/2e \right )I_0$ and $\gamma$ is the normalized dc bias current.
The presence of a bias current $\gamma$ flowing through the junction tilts $U(\varphi,\gamma)$, so that for $\gamma < 1$ the potential shows metastable wells with a barrier height given by~\cite{Ben84}
\begin {equation}\label{eq:barrier}
\Delta\, \mathcal{U}(\gamma)=\frac{\Delta U(\gamma)}{E_{J_0}}=2 \left[ \sqrt{1-\gamma^2} - \gamma \cos^{-1}(\gamma) \right],
\end {equation}
while for $\gamma \geq 1$ the potential profile shows no maxima and minima, see Fig.~\ref{Fig02}(b).
In a current-biased JJ, the left and right potential energy barriers are asymmetric, and therefore, because of the exponential Kramers dependence upon the energy barriers, escapes in the uphill direction are negligible.
Moreover, to achieve a significant, detectable voltage, it is necessary that several of these jumps towards higher energies occur, and therefore the corresponding probability can be safely neglected.
The bias current also changes the shape of the potential wells, and therefore affects the value of the plasma frequency, the small amplitude oscillation  frequency of the particle in the potential well. In the harmonic approximation the frequency can be written as $\omega_p(\gamma)=\omega_Jf_0(\gamma)$, where $f_0 (\gamma) = \left(1-\gamma^2\right)^{1/4}$.

Thermal fluctuations can drive the system out of a washboard minimum. In fact, in the presence of noise the phase solution in the potential minima becomes metastable, and the resulting thermally activated escape rate $r_{\textsc{ta}}(\gamma)$ is given by the Kramers approximation~\cite{Kramers40, Risken89}, for moderate damping, as
\begin{equation}\label{r0_full}
 r_{\textsc{ta}}(\alpha,\gamma,D) = \left( \sqrt{\frac{\alpha^2}{4}+f_0^2(\gamma)}-\frac{\alpha}{2} \right) \frac{e^{-\frac{\Delta\, \mathcal{U}(\gamma)}{ D}}}{2\pi}.
\end {equation}
In the case of a weak dissipation, this equation can be simplified into
\begin{eqnarray}\label{r0}
 r_{\textsc{ta}}(\gamma,D) \simeq \frac{f_0(\gamma)}{2\pi} e^{-\frac{\Delta\, \mathcal{U}(\gamma)}{ D}}.
\end {eqnarray}
In this approximation the escape rate depends only on the noise intensity and the bias current, and allows to identify the current value $\gKr$ at which, for a certain noise intensity $D$, the inverse Kramers rate equals the measurement time, \emph{i.e.}, $r_{\textsc{ta}}^{-1}(\gKr,D)=P$. For current values below $\gKr$,  thermal fluctuations are too weak to trigger escapes before the measurement time $P$. In this condition, and in the absence of external signal, \emph{i.e.}, $\gamma_s=0$, the phase particle remains trapped within the initial metastable well.

Another escape mechanism that can cause the particle to leave the metastable state is the macroscopic quantum tunneling (MQT), whose rate is given by~\cite{Dev85,Bla16}
\begin{equation}\label{MQTescape}
r_{\textsc{mqt}}(\alpha,\gamma) =a_q\frac{f_0(\gamma)}{2\pi}e^{ \left [ -7.2\frac{\Delta U(\gamma)}{\hbar \omega_p(\gamma)}\left ( 1+0.87\alpha \right ) \right ]},\end{equation}
where $a_q=\sqrt{120\pi\left [ \nicefrac{7.2 \Delta U(\gamma)}{\hbar \omega_p(\gamma)} \right ]}$
~\footnote{In our normalized units $\frac{\Delta U(\gamma)}{\hbar \omega_p(\gamma)}\approx\frac{1027}{\alpha R_J}\frac{\Delta\mathcal{U}(\gamma)}{ f_0(\gamma)}$. To draw the $r^{-1}_{\text{MQT}}$ \textit{vs} $\gamma$ curve in Fig.~\ref{Fig07} we set $\alpha=0.025$ and $R_J=1\;\text{k}\Omega$.}.
The tunneling rate is independent of the temperature, being mainly a function of the height $\Delta U$ of the potential barrier to overcome. Thus, decreasing the temperature, the thermal activation rate can reduce so much that macroscopic tunneling processes dominate the escape dynamics.
The temperature that separates the thermal regimes in which the two processes, thermal activation and quantum tunneling, equal each other is called \emph{crossover temperature}, $T_{\text{cr}}$. In the case of $\alpha\ll1$ and $a_q\approx1$, this threshold temperature can be simply estimated as $T_{\text{cr}}\approx\nicefrac{\hbar\omega_p}{7.2k_B}$, and acquires a value of $T_{\text{cr}}\simeq23\;\text{mK}$ if $\omega_p = 22\;\textup{GHz}$. 

In summary, a JJ can be seen as a threshold detector that produces a measurable output (the voltage state) when perturbed by an energy comparable with the energy barrier height given by Eq.~\eqref{eq:barrier}.
The unavoidable intrinsic thermal noise determines random switchings to the voltage state with a certain statistical distribution, which is modified by the absorption of photons.
How the distribution changes, as well as the strategies to distinguish the absorbed photons, is the subject of the following Section.

%
\begin{figure}[t!!]
\includegraphics[width=0.8\columnwidth]{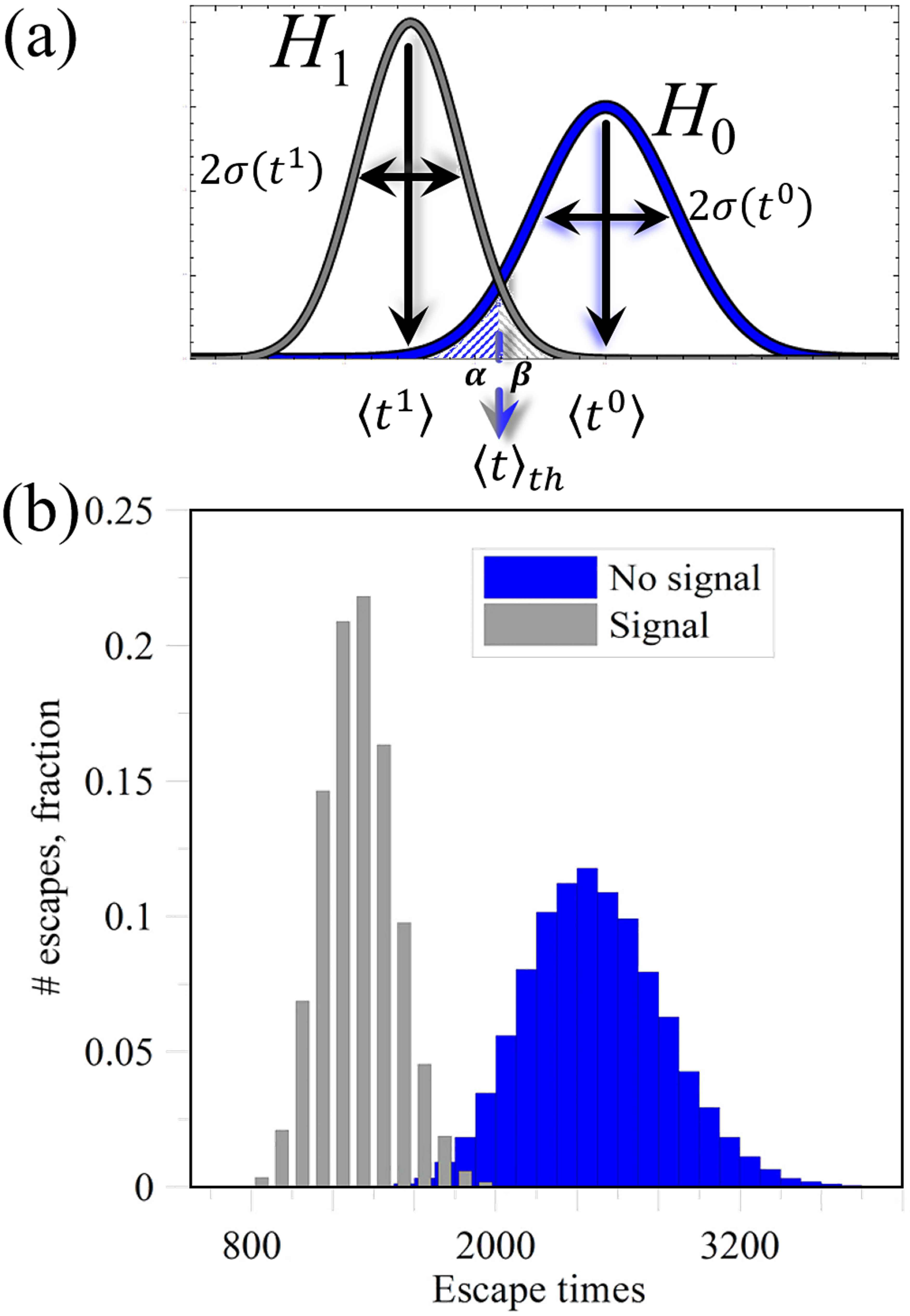}
\caption{(a) Application of the detection scheme to the averaged data, assumed to be Gaussian distributed.
The threshold $\langle t \rangle _{th}$ determines the quantities $\alpha$ and $\beta$, the false detection probability and the probability to miss detection, respectively. (b) Distribution of escape times obtained numerically, with and without the signal and setting the values $D=0.1$, $\gamma_b=0.6$, $A=0.23$, $N=2\times10^6$, $P=10^6$, $\alpha=0.025$, $T=1000$, and $\delta t=10$.}
\label{Fig03}
\end{figure}

\section{The $\dKC$ index}
\label{Sec01-A}

This Section describes a possible approach to quantify the goodness of signal detection~\cite{Lev73,Hel94} in the Josephson framework~\cite{Pie21}.
First, one supposes that the system has been repeatedly prepared, $N$ being the number of such repetitions, in the initial state with the phase particle at rest in the bottom of a washboard well.
Due to external deterministic disturbances and/or thermal fluctuations, during a measurement time $P$, some switches from the superconducting to the voltage state occur.
For each of the $N$ repetitions, we store the time the junction takes to pass from the zero- to the finite-voltage state to construct a distribution of switching times, that can be visualized in the form of a histogram.
Qualitatively, one asks whether this histogram is compatible with the two pivotal hypotheses: the presence of thermal noise alone (hypothesis $H_0$), or the absorption of microwave photons in a thermal noise background (hypothesis $H_1$). This is schematically presented in Fig.~\ref{Fig03}, where the two hypothesis of the binary test are\\

\noindent $H_0 \equiv $ \{Null hypothesis\} $\equiv$ \{Escapes induced by thermal noise alone\}\\

\noindent $H_1 \equiv$ \{Alternative hypothesis\} $\equiv $ \{Escapes induced by thermal noise $+$ the arrival of current pulses at a rate $ r_A = \nicefrac{1}{T } $\}.
\\

\noindent The blue curve on the right in Fig.~\ref{Fig03}(a) indicates the distribution under the null hypothesis $H_0$, while the gray curve to the left is the distribution under the alternative hypothesis $H_1$. 
Analogously, in panel (b) we show two distributions of switching times obtained numerically for a specific combination of system parameters, with and without the exciting signal.

 The chosen value of the threshold $\langle t \rangle _{th}$ determines the quantities $\alpha$ and $\beta$ highlighted in Fig.~\ref{Fig03}.
In detail, the parameters $\alpha$ and $\beta$ indicate the false detection probability, \emph{i.e.}, that the switching has been caused by a purely thermal activated escape and the probability that the signal is missed, respectively.
Therefore, the lower the quantities $\alpha$ and $\beta$, the better the test.
If the two distributions are well separated, it is possible to choose an appropriated threshold $\langle t \rangle_{th}$ that allows both a low significance, \textit{e.g.}, $\alpha \leq 1\%$ to exclude that the pure thermal noise explains the results, and a sufficiently high test power, \textit{e.g.}, $1-\beta \geq 99\%$ to guarantee that the presence of a signal has not gone amiss.
We underline that the values of $\alpha=\beta=1\%$ are purely indicative: in practice, the acceptable values for $\alpha$ and $\beta$ depend on the specific application. For instance, in the case of rare particles experiments, where indications about the possible presence of a particle is important, even quite high $\alpha$ and $\beta$ values could be acceptable. Conversely, in telecommunication, where the adequate error rate is very small, the values of $\alpha$ and $\beta$ are typically much lower.

A convenient way to quantify the discrimination power of a detection strategy is the KC index $\dKC$~\cite{Kumar84}. If the switching occurs in all the $N$ experiments, the $\dKC$ index can be defined as
\begin{equation}
\label{kcindex}
d_ {\textsc{kc}}= \frac{\left | \langle\tau_{\textsc{sw}}\rangle_{1} - \langle\tau_{\textsc{sw}}\rangle_{0} \right |}{\sqrt{\frac{1}{2}\left[\sigma^2(\tau_{\textsc{sw}})_{1} + \sigma^2(\tau_{\textsc{sw}})_{0}\right]}},
\end{equation}
where
\begin{equation}\label{average_tau}
\langle \tau_{\textsc{sw}} \rangle_1 = \frac{1}{N}\left . \sum_{i=1}^N \tau_i \right |_{1}
\end{equation}
is the estimated average value of the switching times in the presence of the signal and noise, and
\begin{eqnarray}\label{variance_mean}\nonumber
\sigma^2(\tau_{\textsc{sw}})_{1} &=& \text{Var}\Big [ \langle \tau_{\textsc{sw}} \rangle_1\Big ] =\text{Var}\left [\left .\frac{\tau_1+\tau_2+...+\tau_N}{N} \right |_{1}\right ] \\
&=& \frac{1}{N \left( N-1 \right)}\sum_{i=1}^N\Big( \tau_i - \langle \tau_{\textsc{sw}} \rangle _1 \Big)^2
\end{eqnarray}
is the corresponding estimate of the variance of the average switching time.
The same quantities with the subscript ``${0}$'' describe the case of the absence of the signal and presence of noise alone.
The KC index defined in Eq.~\eqref{kcindex} represents a qualitative indication, depending on the specific statistical distribution, and corresponds to the SNR only in the case of Gaussian distributions.
However, one can use the $\dKC$ index as a first estimate of the capability to discriminate the two conditions (with or without photon-induced switches).
In the framework of microwave field detection, a quite similar statistical estimator, called Hellinger distance~\cite{Spe14}, is used to characterize distributions of switching currents in a JJ operating as a qubit~\cite{Oel17}.

Indeed, by comparing the average in Eq.~\eqref{average_tau} (more accurately, the sample mean) with the corresponding average in the absence of external signal, the index is clearly related to the $t$-test statistics. Although this is a good indicator, it is clear, however, that a full test is needed to compare the entire information content of the escape times distribution and, therefore, usually can outperform the $t$-test~\cite{Addesso12}.
The $\dKC$ index has been already employed for a Josephson-based detection scheme in the case, however, of a sinusoidal exciting signal, for both a fixed~\cite{Filatrella10,Addesso12} and a linearly ramping bias current~\cite{Pou20}.

The strategy presented above exploits the information content of the switching time distribution.
This strategy is only possible if during each experimental run, that we suppose of duration $P$, see Fig. \ref{Fig02}(a), a switching occurs.
If this is not the case, that is, if in at least in one of the $N$ pulse trains no switchings occur, it is not possible to determine the escape time in all the $N$ experiments, and at least one of the terms in Eqs. (\ref{average_tau},\ref{variance_mean}) remains undetermined.
In fact, Eq.~\eqref{kcindex} is formulated assuming that all switching events occur within the observation time $P$.
Conversely, if a portion of the measurements does not exhibit switching events during the time $P$, we expect that the distribution of collected switching times might be relatively poor and somehow deformed, since a portion of the switching times, those longer than $P$, are not recorded. When this is the case, one has only a limited information on the escape times, and can for instance assume that the average time reads exactly $P$, and the standard deviation, consequently, vanishes, $\sigma\left( \tau_{sw}\right) =0$.
However, this is but an estimate; more accurately one can just record whether or not a switching event has occurred during the measurement time.
We thus propose to determine the information content of the experiments in which the escape time passes the maximum waiting time $P$ with an alternative definition of the KC index.
It is in fact conceivable to detect the signal in a thermal noise background by grouping the measurements in two classes: those in which at least one escape process occurs within the measurement time $P$, and those in which it is not.
This \emph{dichotomous} classification of the experimental runs can be dealt with a counting statistics.
To do so, one defines the probabilities, $p_0$ and $p_1$, of the occurrence of a switching prior to the measurement time $P$ under the two hypotheses $H_0$ and $H_1$, respectively, as
\begin{equation}\label{pswitch_p0}
p_0 = {\mathcal P}\left( \tau_i \le P \lvert H_0 \right) \quad\text{and}\quad p_1 = \mathcal{P}\left( \tau_i \le P \lvert H_1 \right),
\end{equation}
which have to be estimated from the experiments.
In the case of a finite $N$, the estimated probabilities read
\begin{subequations}\label{p_estimate}
\begin{align}\label{p0_estimate}
&\widetilde{p}_0 \simeq \frac{N_0}{N}, \quad N_0 \equiv \text{number of } \tau_i \le P \lvert H_0 \\
\label{p1_estimate}
&\widetilde{p}_1 \simeq \frac{N_1}{N}, \quad N_1 \equiv \text{number of } \tau_i \le P \lvert H_1 .
\end{align}
\end{subequations}
The estimates are Gaussian distributed with a variance $\sim\widetilde{p}_j\left(1-\widetilde{p}_j\right)$, with $j=1$ or $0$, so that one can evaluate the SNR through the \emph{binomial} KC index, $\tdKC$, defined as
 \begin{equation}
 \label{kcindex_p}
\tdKC = \sqrt{N} \frac{|\widetilde {p}_1 - \widetilde{p}_0|}
 {\sqrt{\frac{1}{2} \left[
 \widetilde{p}_1\left(1-\widetilde{p}_1\right) + \widetilde{p}_0 \left(1-\widetilde{p}_0\right)
 \right]}}
\end{equation}
if at least one of $\widetilde {p}_0$ and $\widetilde {p}_1$ is not equal to $0$ or $1$, and $\tdKC =0$ otherwise.
If a switching occurs in any run, the probabilities in Eq.~\eqref{p_estimate} assume the value $\widetilde{p}_j = 1$, so that the SNR with this method vanishes.
In the opposite case, that is if no switching occurs in the experiments, one obtains $\widetilde{p}_j=0$, that also gives a vanishing SNR.
Therefore, a detection method based on the dichotomous grouping of the switching events (\ref{pswitch_p0}) has to be adopted only when the switching occurs for some, but not all, the $N$ experiments~\cite{Fil21}.
A further distinction has to be made. In the limit case in which in one condition all the escape times occur prior to the limit time $P$, say $p_1=1$, and the other distribution consists of switching time longer than the limit $P$, $p_0=0$, Eq.(\ref{kcindex_p}) is not defined as the denominator vanishes.
When this is the case we assume $\tdKC = 0$. However, the detection is likely to be very good, for the two responses are well distinct: no switchings without signal, and switchings in every run if the signal is there.

\section{Numerical Details}
\label{Sec03}\vskip-0.2cm
Equation~\eqref{eqJJ_norm} is simulated by means of an Euler algorithm, with a small enough time step $h=5\times 10^{-3}$. We use the Box-Mueller algorithm to generate a Gaussian distribute noise, $\zeta = \sqrt{-4Dh\log(r_{1})}\cos(2 \pi r_{2}),$ from two random numbers, $r_{1}$ and $r_{2}$, which are uniformly distributed over the unit interval~\cite{Fox88}.

To compute the mean switching times $\tau_i$ and the standard deviation $\sigma$, we integrate the stochastic equation~\eqref{eqJJ_norm} and results are averaged over $N$ independent repetitions (typically, $N\in[10^{3}-10^{4}]$). During each numerical experiment, we record the time at which the Josephson phase particle $\varphi$ leaves the potential well in which it resides, \emph{i.e.}, the instant at which $\varphi\textgreater\pi-\arcsin(\gamma_{b})$.
The maximum integration time used in this work is $P=10^{6}$.
In the initial conditions the phase particle is at rest in the bottom of the potential well, that is, $\varphi(0)=\arcsin(\gamma_{b})$ and $\varphi'(0)=0$, \emph{i.e.}, in the zero-voltage state.
The time $\langle \tau_{\textsc{sw}} \rangle$ thus calculated is a mean first passage time: when the phase particle overcomes the threshold, the simulation is cut short and the system is prepared again in the prescribed initial state. 
We stress that the numerical simulations returns the full collection of switching times in all the $N$ independent numerical repetitions, for each combination of the parameter values. From these collected data we calculate first the average switching time and the standard deviations, and then the $\dKC$ and $\tdKC$ indexes.

A word of caution: in a real experiment an additional time to set the initial conditions after each passage is required to restore the initial state. This additional time should be included in a careful analysis of the performances of the signal detection. However, for the sake of simplicity, we consider only the escape time in our scheme.

\section{Results and Discussions}
\label{Sec02}\vskip-0.2cm

In the following, the performances of the detection method are analyzed in terms of the intensity of the signal (that is, the pulse amplitude $A$), the bias current value $\gamma_b$, and the temperature of the system (that is, the noise intensity $D$).
As the system is subject to a low damping and operates at low temperatures, the dissipation parameter is set to $\alpha=0.025$, and two different intensities of thermal noise, $D=\{0.01, 0.1\}$ are considered.
The simulation is repeated for different values of the amplitude $A$ of the driving periodical signal.
The distance between two consecutive pulses is $T=1000$, and the width of each pulse is $\delta t=10$. See Fig. \ref{Fig02} for the meaning of these parameters.

\begin{figure}[t!!]
\includegraphics[width=0.5\textwidth]{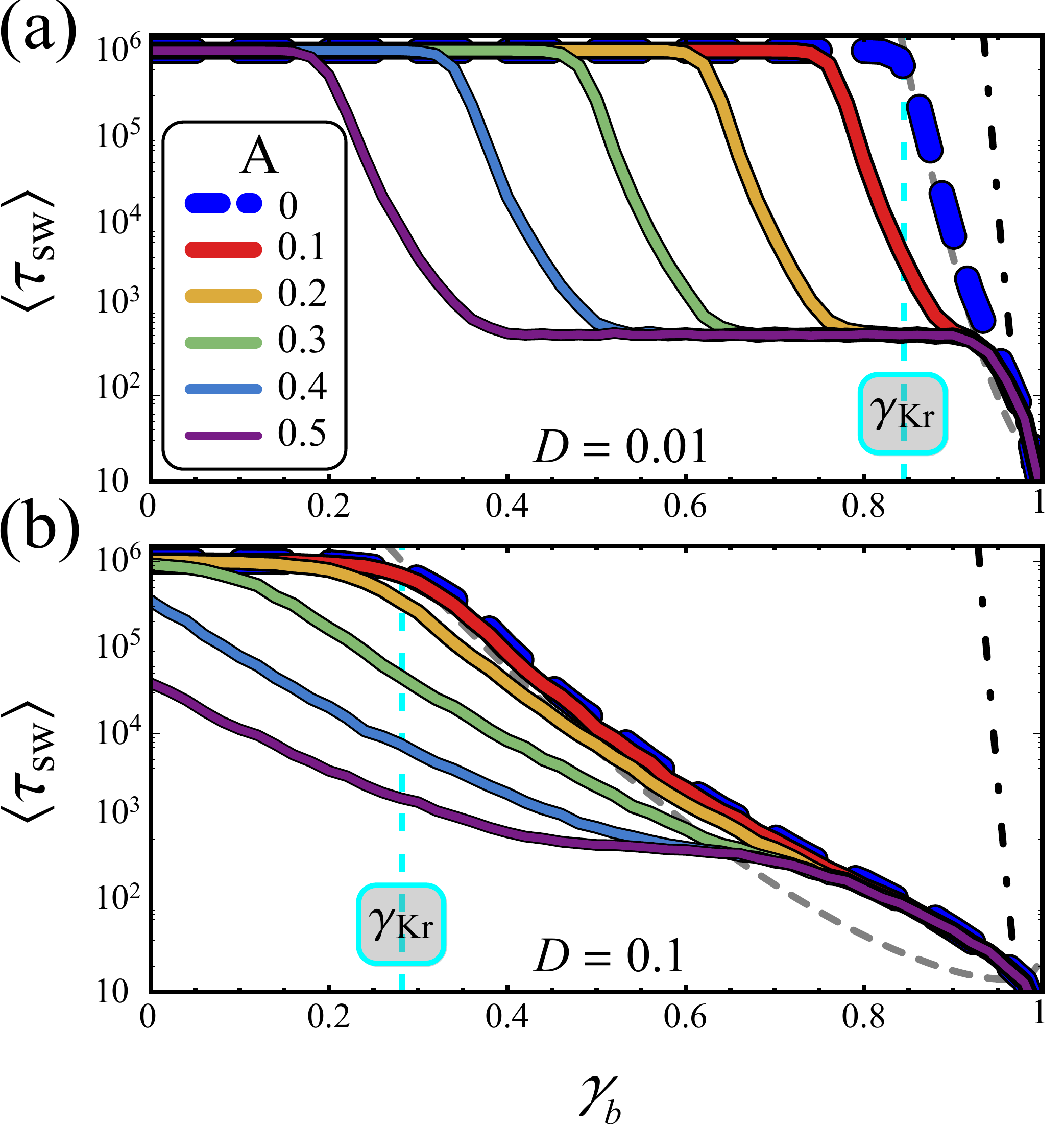}
\caption{Mean switching time, $\tSW$, as a function of the normalized bias current $\gamma_b$ at different values of $A$, in the case of $D=0.01$ (a) and $D=0.1$ (b). The blue dashed curve indicates the average switching time in the case of purely noise-activated escape processes. The values of the other parameters are: $N=10^3$, $P=10^6$, $\alpha=0.025$, $T=1000$, and $\delta t=10$. The gray dashed and black dot-dashed lines indicate the inverse Kramers and MQT rates, $r^{-1}_{\textsc{ta}}(\alpha,\gamma_b,D)$ and $r^{-1}_{\textsc{mqt}}(\alpha,\gamma_b)$, calculated according to Eqs.~\eqref{r0_full} and~\eqref{MQTescape}, respectively. The vertical cyan dashed line marks the noise amplitude, $\gKr$, at which $r^{-1}_{\textsc{ta}}(\alpha,\gKr,D)=P$. The legend in (a) refers to both panels.}
\label{Fig04}
\end{figure}

The detector is investigated through the obtained values of the average switching time $\tSW$, and of the KC indexes $\dKC$ and $\tdKC$.
As expected, the SNR changes with the JJ parameters even if the amplitude $A$ of the pulses and the temperature are fixed.
Among the JJ parameters, the bias current is presumably the easiest to tune, it is therefore interesting to search for its the best values to detect a signal.
Even without the driving signal, \emph{i.e.}, $\gamma_s=0$, or equivalently $A=0$, thermal fluctuations can push the phase particle out of the metastable well. The presence of a signal, $\gamma_s\neq0$, is expected to further facilitate the escape process, that is to reduce $\tSW$.
This is shown in Fig.~\ref{Fig04}, that displays the bias-current-dependence of average switching time in the case of a dynamics determined by thermal noise alone ($A=0$, denoted with $\tSW_0$), and for increasing pulse amplitudes ($A\in[0.1-0.5]$, denoted with $\tSW_{\textsc{a}}$).
Figures~\ref{Fig04}(a) and \ref{Fig04}(b) correspond to two different noise intensities, $D=0.01$ and $D=0.1$, respectively, while the other parameter values are kept fixed to $N=10^3$, $P=10^6$, $\alpha=0.025$, $T=1000$, and $\delta t=10$.
A vertical cyan dashed line indicates the threshold bias current, $\gKr$, above which Kramers theory predicts that thermal fluctuations, corresponding to the noise intensity $D$, trigger escape events within the measurement time $P$.
In fact, thermally activated switchings can only occur if the potential barrier is sufficiently low, that is, if the bias current tilts enough the potential $U$.
In the case of switchings induced by noise alone, the average switching time $\tSW_0$ saturates at  $\tSW_0\simeq P$ for $\gamma_b\lesssim\gKr$, see the blue dashed curve. This corresponds to the fact that, for low $\gamma_b$ values, thermal noise alone is not able to cause escape processes within the measurement time $P$.
Conversely, for higher $\gamma_b$ values the potential barrier is low enough that sizable thermally induced switchings are permitted.
To estimate the system response in this regime, we include in Fig.~\ref{Fig04}, as a gray dashed curve, the Kramers time, $r^{-1}_{\textsc{ta}}(\gamma_b)$, as per Eq.~\eqref{r0_full}.
The mean switching time well agrees with the estimate $r^{-1}_{\textsc{ta}}(\gamma_b)$ behavior above $\gKr$, while it diverges from this estimate as $\gamma_b$ increases.
In other words, the numerical results are poorly described by Kramers theory in the high-current regime, that is, when the height of the potential barrier is quite small. This is not surprising, inasmuch the Kramers approach describes the evolution from a metastable state in the case of a noise intensity much smaller than the potential barrier height.

%


For completeness, we insert in Fig.~\ref{Fig04} also a black dot-dashed curve tracing the inverse MQT rates, $r^{-1}_{\textsc{mqt}}(\gamma_b)$, calculated according to Eq.~\eqref{MQTescape}.
In the temperature regimes explored in this work, $r^{-1}_{\textsc{mqt}}(\gamma_b)$ is much larger than the other characteristic times.
This means that macroscopic quantum tunneling processes do not play a role and only thermally activated switching dynamics are to be taken into account.

\begin{figure*}[t!!]
\includegraphics[width=\textwidth]{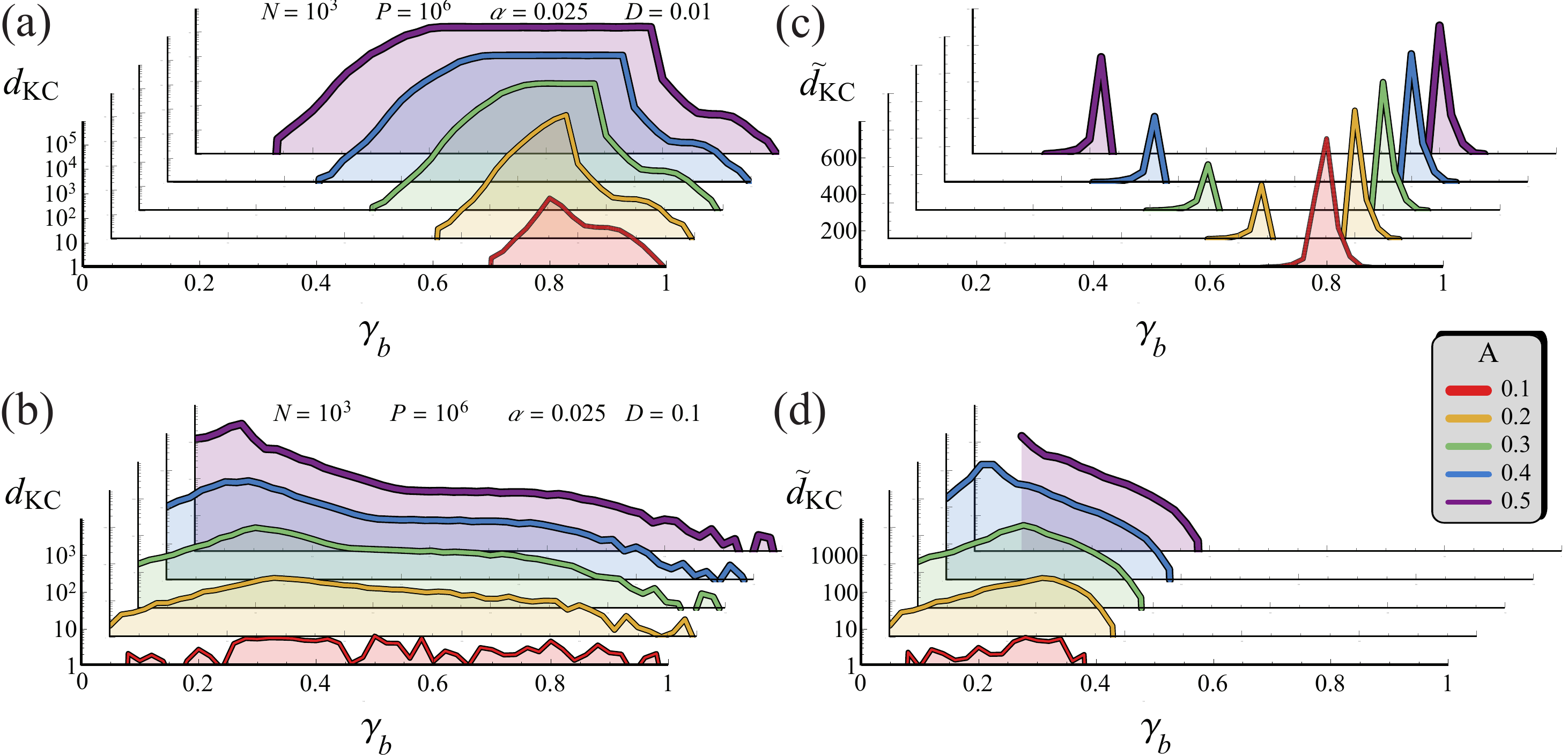}
\caption{Conventional and binomial KC indexes, $\dKC$ and $\tdKC$, as a function of the normalized bias current $\gamma_b$ at different values of $A$, in the case of $D=0.01$ [top panels (a) and (c)] and $D=0.1$ [bottom panels (b) and (d)]. The values of the other parameters are: $N=10^3$, $P=10^6$, $\alpha=0.025$, $T=1000$, and $\delta t=10$. The legend in panel (d) refers to all panels.}
\label{Fig05}
\end{figure*}

In more details, Fig.~\ref{Fig04}(a) displays the mean switching time in the low noise intensity case, $D=0.01$.
The blue-dashed curve for $A=0$ exhibits a plateau $\tSW_0\simeq P$ for $\gamma_b$ values below the threshold value $\gKr\simeq0.84$, and an exponential decrease for $\gamma_b>\gKr$.
For $A>0$ the behavior of the $\tSW_{\textsc{a}}$ \emph{vs} $\gamma_b$ curves is more complex, \emph{i.e.}, four different response regimes can be distinguished.
At low bias currents, $\gamma_b\lesssim \left ( \gKr-A \right )$, it appears a plateau at $\tSW_{\textsc{a}}\simeq P$ where no switching events occurs within the measurement time $P$, despite the action of both thermal noise and impulsive current.
With a further increase of the bias current, the $\tSW_{\textsc{a}}$ \emph{vs} $\gamma_b$ curves tend to decrease exponentially.
In this range of $\gamma_b$ values the current pulses induce, with a certain probability, a voltage switch.
The width of this current values range depends on the noise intensity $D$, and vanishes for $D\to0$, for the transition is abrupt in the absence of noise.
Moreover, the reduction of the mean switching time indicates that the switching probability grows as the bias current increases, up to the $\gamma_b$ point where the first current pulse alone triggers the escape from the metastable state.
From this point on, the phase dynamics at higher $\gamma_b$ values does not change, \emph{i.e.}, the potential is so tilted that the first pulse induces an escape, so that $\tSW_{\textsc{a}}$ tends to a constant value, which is independent of $\gamma_b$ and dominated by the arrival rate of the first pulse.
Of course, the greater the value of $A$, the larger the range of $\gamma_b$ values that exhibits the lower plateau at $\tSW_{\textsc{a}}\simeq500$, see Fig.~\ref{Fig04}(a).
The actual $\tSW_{\textsc{a}}$ value only depends on the randomly distributed initial waiting time $T_0$, \emph{i.e.}, on the time the first pulse takes to arrive in the range $[0-T]$; in the present case the distance between two consecutive pulses reads $T=1000$.
Finally, for very high bias currents, $\gamma_b\gtrsim0.92$, all curves coalesce, and $\tSW_{\textsc{a}}$ is drastically reduced for higher $\gamma_b$ values.
In this regime, the potential barrier is so small that the system switches to the voltage state before the arrival of the first pulse.
As the initial waiting time is $T_0 \leq T$, the current value $\widetilde{\gamma}_{\text{Kr}}$ after which the dynamics is dominated by the noise can be roughly estimated as $r_{\textsc{ta}}^{-1}(\widetilde{\gamma}_{\text{Kr}},D)=T$, that is $\widetilde{\gamma}_{\text{Kr}}\approx0.92$ for $T=1000$ and $D=0.01$.

In Fig.~\ref{Fig04}(a) a neat separation between the curves with and without signal occurs, as we have motivated above, around the current value $\gamma_{\text{Kr}}$.
If the noise intensity is increased to the value $D=0.1$, see Fig.~\ref{Fig04}(b), the threshold bias current is lowered to $\gKr\simeq0.28$ and the $\tSW$ \emph{vs} $\gamma_b$ curves are closer to each other and less separated by the unperturbed, purely thermal curve $\tSW_0$ (that deviates from the $r^{-1}_{\textsc{ta}}(\gamma_b)$ Kramers theory, as expected for high noise intensity).
For instance, the curve $\tSW_{\textsc{a}}$ for $A=0.1$ tends to the pure-noise curve $\tSW_0$ for almost any value of the bias.
In general, all $\tSW_{\textsc{a}}$ curves for $A\in[0.1-0.5]$ superimpose at high current values, \emph{i.e.}, $\gamma_b\gtrsim0.75$.
Under these circumstances, in which the data obtained in different conditions roughly overlap, only a careful analysis of the switching times can distinguish whether the impulsive signal is present or not.

It is important at this point to interpret the results in terms of the possibility to detect current pulses. Let us recall that we consider the JJ as the detector, and that, through the analysis of the average escape time $\tSW$, it should be possible to decide if the pulses have been absorbed.
It is therefore convenient that the undisturbed escape time curve, the dashed line in Fig.~\ref{Fig04}, is  separated as much as possible from the curves obtained in the presence of pulses.
In the low-noise case, $D=0.01$, shown in Fig.~\ref{Fig04}(a) the signal analysis is straightforward, being the $\tSW$ curves well separated for almost all $\gamma_b$ values.
In other words, at a given $\gamma_b\gtrsim(\gKr-A)$ there is an indication of how easy would be to detect the presence of an external signal embedded in a thermal noise background.
If the curves are well separated, and enough data are provided ($N=10^3$ in the numerical examples), by inspection of the mean switching time distributions it is possible to decide.
A quite different scenario emerges at high bias currents, $\gamma_b\gtrsim\gKr$, where all $\tSW_{\textsc{a}}$ curves tend to coalesce. Moreover, if one assumes a much smaller signal amplitude, \emph{i.e.}, $A\ll0.1$, the $\tSW_{\textsc{a}}$ and $\tSW_0$ curves are superimposed and the analysis is problematic.
Similarly, at higher noise intensity, \emph{e.g.}, $D=0.1$, one cannot rely only on a simple analysis of the average switching times, especially for the lower signal amplitude, $A\simeq 0.1$, because the $\tSW$ \emph{vs} $\gamma_b$ data in the pure-noise and pulse-sustained experiments overlap appreciably, see the blue-dashed and the red curves in Fig.~\ref{Fig04}(b).
To make quantitative these observations, it is essential to use an estimator of the goodness of detection such as the $\dKC$ index.

In Fig.~\ref{Fig05} we show both the conventional and the binomial KC indexes, $\dKC$ and $\tdKC$, as a function of $\gamma_b$, calculated through Eqs.~\eqref{kcindex} and~\eqref{kcindex_p}, respectively.
Here, we choose the parameter values as in Fig.~\ref{Fig04}. In particular, the top panels of this figure are obtained for $D=0.01$, while the bottom panels for $D=0.1$.

We start discussing the $\dKC$ \emph{vs} $\gamma_b$ behavior in the low-noise case, see Fig.~\ref{Fig05}(a) for $D=0.01$. As expected, the KC index increases with the amplitude $A$ of the signal, while the different $\dKC$ curves, obtained by changing $A$, look quite similar. In particular, at low bias currents, \emph{i.e.}, $\gamma_b\lesssim(\gKr-A)$, the $\dKC$ vanishes, since the mean switching times are almost independent of the signal amplitude, being $\tSW_0\simeq\tSW_{\textsc{a}}\simeq P$. This suggests that this region of $\gamma_b$ values is not suitable for detection. Instead, increasing $\gamma_b$, we observe the upsurge of the KC index, since $\tSW_{\textsc{a}}\simeq500\ll\tSW_0\simeq P$. In correspondence of the $\gamma_b$ values giving the plateau at $\tSW_{\textsc{a}}\simeq500$ shown in Fig.~\ref{Fig04}(a) we obtain a similar plateau with a $\dKC\gg1$. Finally, at even larger bias currents, $\gamma_b\gtrsim\gKr$, we observe that the $\dKC$ tends first to the value $\dKC\simeq40$, but then it reduces drastically when $\gamma_b\to1$, for all pulse amplitudes.

\begin{figure}[t!!]
\includegraphics[width=\columnwidth]{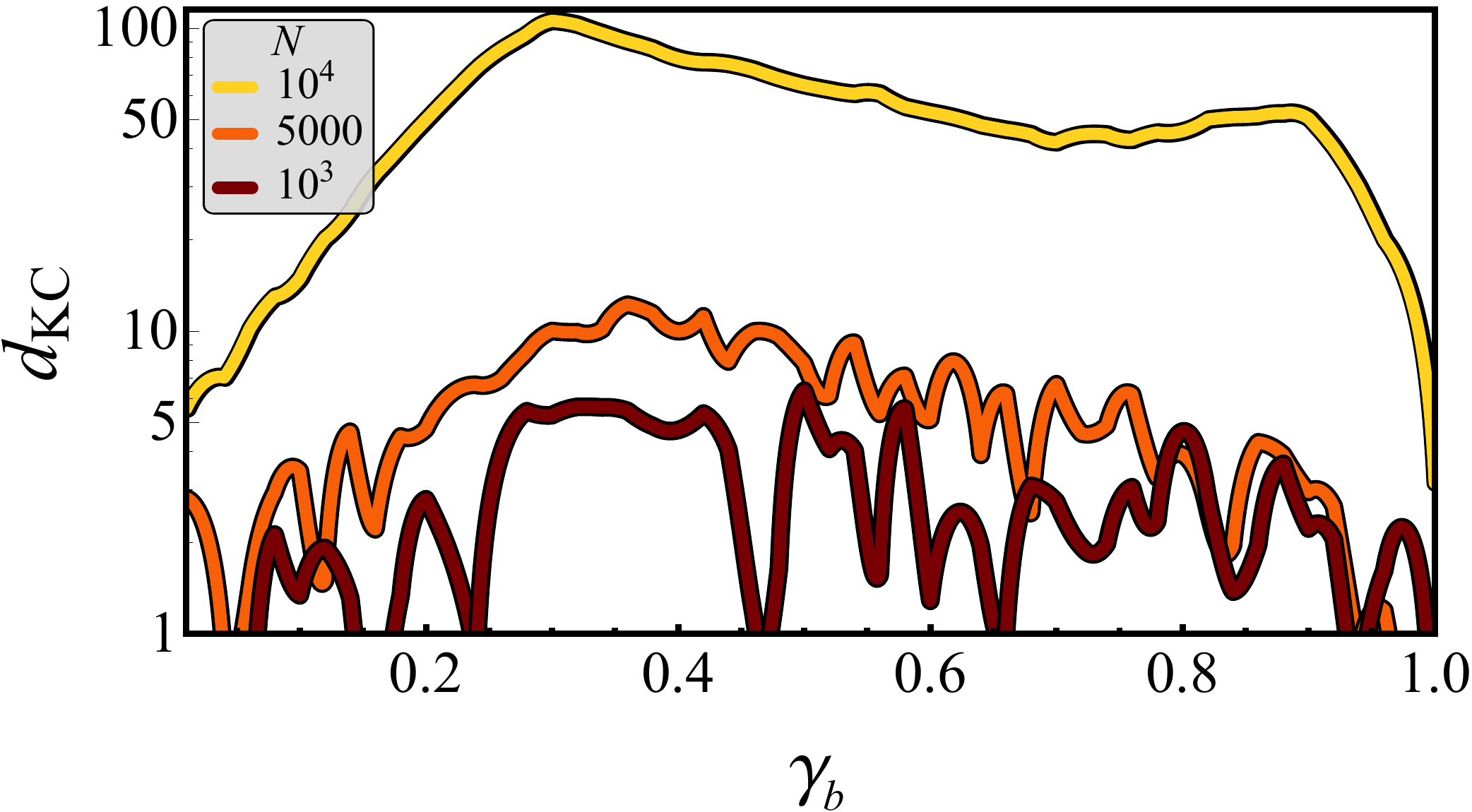}
\caption{KC index, $\dKC$, as a function of $\gamma_b$, at different values of $N$. The values of the other parameters are: $D=0.1$, $A=0.1$, $P=10^6$, $\alpha=0.025$, $T=1000$, and $\delta t=10$.}
\label{Fig06}
\end{figure}

Increasing the noise intensity, the $\dKC$ \emph{vs} $\gamma_b$ behavior changes, as shown in Fig.~\ref{Fig05}(b) for $D=0.1$. In particular, for $A>0.1$ a quite large $\dKC\gg1$ non-monotonic behavior appears, that exhibits a maximum at $\gamma_b\approx\{0.28, 0.20, 0.14, \text{ and } 0.08\}$ for $A=\{0.2,0.3,0.4, \text{ and } 0.5\}$, respectively.
Instead, in the case of $A=0.1$, the $\dKC$ \emph{vs} $\gamma_b$ curve is quite scattered, notwithstanding the existence of a large interesting region of $\gamma_b$ values giving a $\dKC\approx5$.
This means that, even if the mean switching times in the pulse-sustained and in the pure-noise cases are almost indistinguishable, see Fig.~\ref{Fig04}(b), a signal with $A=0.1$ embedded in a thermal noise background with fluctuations intensity $D=0.1$ is still detectable if a proper level of bias current is used.
Moreover, the detector performances can be improved increasing the number of experiments $N$. This is shown in Fig.~\ref{Fig06} for $D=0.1$ and $A=0.1$, comparing the results for the cases $N=10^3$ and $N=10^4$ repetitions.

The $\dKC$ \emph{vs} $\gamma_b$ curve in light color for $N=10^4$ shows a non-monotonic behavior, characterized by two distinct maxima, \emph{i.e.}, $\dKC\approx106$ at $\gamma_b=0.3$ and $\dKC\approx52$ at $\gamma_b=0.88$.
These are the ideal low- and high-current conditions to perform the detection.
In other words, one can retrieve the information on the presence of the pulses in two working regimes, \emph{i.e.}, two quite different levels of potential energy barrier.
In the two cases the typical switching times are quite diverse. In fact, at high bias currents the escape dynamics is very fast even in the pure-noise case, while at low bias currents the average switching time is expected to be quite long, \emph{i.e.}, $\tSW\gtrsim P$ with $P=10^6$. To briefly summarize this part, one can achieve the desired level of the KC index increasing the number of experiments $N$ and optimizing the bias point.

In Fig.~\ref{Fig05} is presented the behavior of the binomial KC index, $\tdKC$ in Eq.~\eqref{kcindex_p}, as a function of $\gamma_b$ at different signal intensities, $A\in[0.1-0.5]$, and for two different noise intensities, $D=0.01$ and $D=0.1$, see panels (c) and (b), respectively.
This estimator of the SNR is different from zero if in one process (either the pure-noise or the pulse-sustained case) escapes occur in some, but not all, the numerical experiments.
In the case of $D=0.01$, when $A>0.1$ the $\tdKC$ \emph{vs} $\gamma_b$ is double-peaked, while for $A=0.1$, the two peaks are close enough to appear as a single peak, see Fig.~\ref{Fig05}(c).
In particular, the low-current peak indicates the situations where pure-noise switches never occur within the measurement time $P$, while the pulse-sustained switches take place with a finite probability.
Conversely, the high-current peaks indicate the situation where pure-noise switches occur with a finite probability, while the pulse-sustained switches certainly happen in all experiments.
This is why the position of the latter peak is independent of the pulse intensity $A$.
At $\gamma_b$ values between the two peaks, the SNR estimate vanishes, $\tdKC=0$, for one never observes pure-noise switches, while the pulse-sustained switches occur systematically.
Finally, for $\gamma_b>0.86$ (or for $\gamma_b\to0$) one observes $\tdKC=0$, for both the pure-noise and the pulse-sustained switching occur (or do not occur) in every experiment.

The $\tdKC$ \emph{vs} $\gamma_b$ behavior changes at higher value of the noise intensity, $D=0.1$, see Fig.~\ref{Fig05}(d).
First for $\gamma_b\gtrsim0.38$, the SNR estimate reads $\tdKC=0$, for both the pure-noise and the pulse-sustained conditions result in escapes within the measurement time $P$ for all runs.
At lower $\gamma_b$, the $\tdKC$ behaves non-monotonically.
Interestingly, for $\gamma_b<0.08$ both the $A=0.1$ and $A=0.5$ cases show $\tdKC=0$, but for different reasons.
In fact, within the measurement time $P$, for $A=0.1$ at low bias current both the pure-noise and the pulse-sustained case show no switchings, while for $A=0.5$ we have no pure-noise-induced escapes although the pulse-sustained dynamics show switching events in all experiments.
In the latter case $\tdKC$ is not defined [see Eq.\eqref{kcindex_p}], but the detection is presumably very effective, see the discussion at the end of Section~\ref{Sec02}.

\begin{figure}[t!!]
\includegraphics[width=\columnwidth]{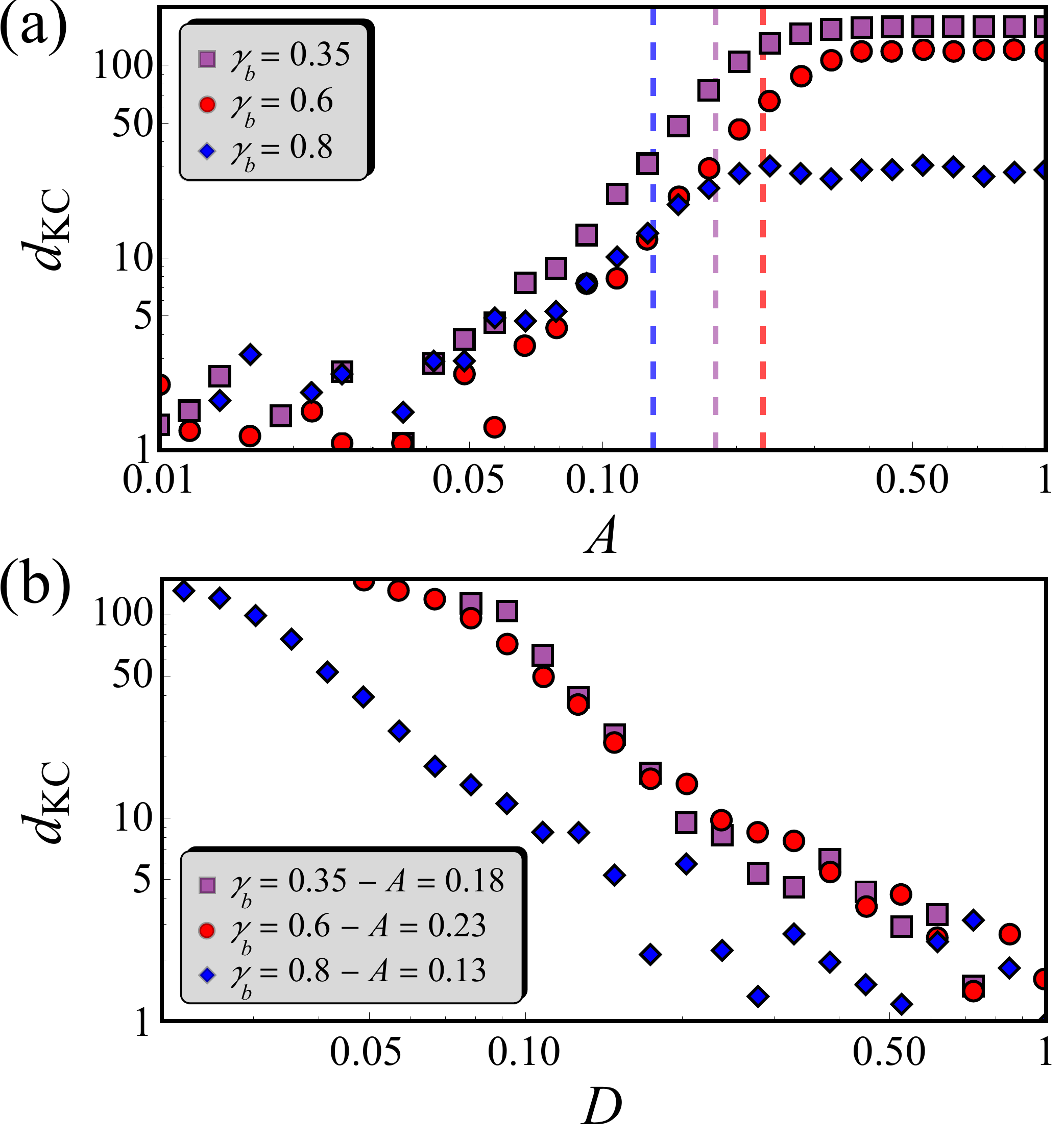}
\caption{KC index, $\dKC$, as a function of $A$, at different values of $\gamma_b$ and $D=0.1$, and $D$, at different values of $\gamma_b$ and $A$, see panels (a) and (b), respectively. The vertical dashed lines in panel (a) indicate the $A$ values chosen to draw the curves in panel (b). The values of the other parameters are: $N=10^4$, $P=10^6$, $\alpha=0.025$, $T=1000$, and $\delta t=10$.}
\label{Fig07}
\end{figure}

Finally, Fig.~\ref{Fig07}(a) shows how the KC index depends on the pulse amplitude $A$ at fixed $D$, and conversely Fig.~\ref{Fig07}(b) displays the dependence of the index as a function of noise at fixed bias points and amplitudes.
In particular, in panel (a) we have a fixed noise intensity $D=0.1$ and different $\gamma_b=\{0.35,0.6,\text{ and }0.8\}$.
All curves exhibit an S-shaped envelop; for $A\lesssim0.05$ the $\dKC$ values are quite low and highly scattered, therefore this region of $A$ values is not recommended for signal detection.
Instead, for higher signal amplitudes the $\dKC$ exponentially increases, up to a threshold $A$ value, after which the $\dKC$ saturates, i.e., it is independent on the amplitude $A$; however, the $\dKC$ saturation value depends on $\gamma_b$.
Thus, the detection is favored by a sufficiently large-amplitude signal, although, depending on the bias current value, the specific performances, \emph{i.e.}, the KC index value, do not change appreciably above a certain amplitude value.

In Fig.~\ref{Fig07}(a), we use vertical dashed lines to evidence the signal amplitudes $A$ at which the $\dKC$ takes a value roughly equal to half of the maximum value reached in the high-amplitude plateau. We set just these values of $A$ to show how the KC index depends on the noise intensity, \emph{i.e.}, to trace the $\dKC$ \emph{vs} $D$ curves shown in Fig.~\ref{Fig07}(b). This figure reveals that the $\dKC$ exponentially decreases by increasing $D$. Moreover, we observe that the curve at high currents, \emph{i.e.}, $\gamma_b=0.8$, is shifted towards low noise amplitudes, with respect to the other two curves at $\gamma_b=0.6$ and $0.35$, which seem instead to overlap.
In other words, referring to the values in Fig.~\ref{Fig07}(b), to obtain the same level of $\dKC$, if $\gamma_b=0.8$ the noise amplitude must be approximately half of that necessary if $\gamma_b=0.6$ or $0.35$. Summing up, Fig.~\ref{Fig07}(b) tells us that, at a given $\gamma_b$, a higher noise amplitude $D$ negatively affects the detection performances. This is due to the fact that by increasing $D$ the mean switching time distribution in the pure-noise case, $\tSW_0$, tends to reduce drastically, \emph{e.g.}, compare the dashed curves in Figs.~\ref{Fig04}(a) and (b), getting closer and closer to the $\tSW_{\textsc{a}}$ distribution in the pulse-sustained case. This confirms the expected result that reducing the temperature is an effective way to improve the detection performances, at least as long as MQT effects do not enter in play.

\begin{figure}[t!!]
\includegraphics[width=\columnwidth]{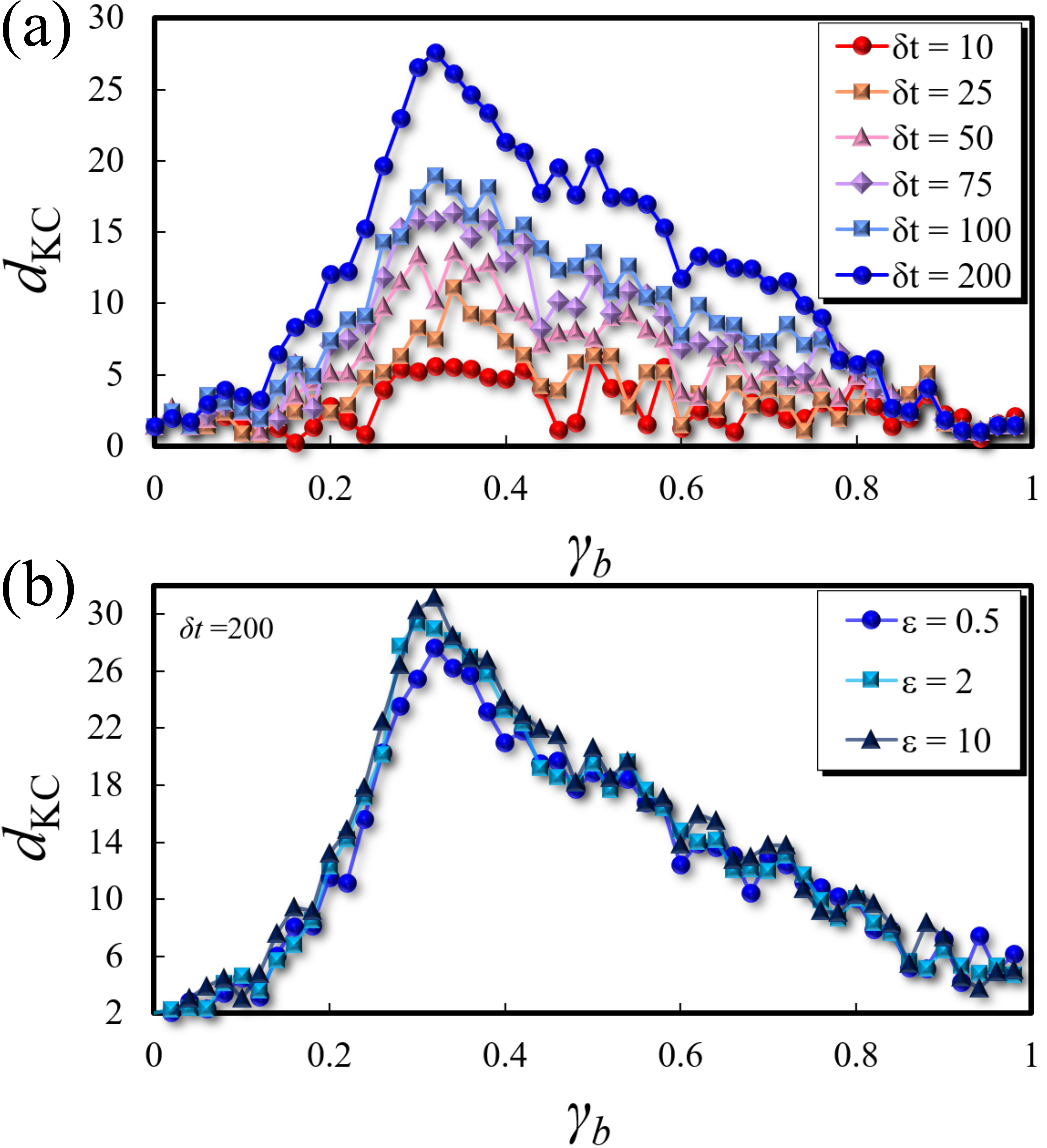}
\caption{KC index, $\dKC$, as a function of $\gamma_b$, at (a) different values of $\delta t$ and (b) different values of $\epsilon$ and $\delta t=200$ in the case of rounded current pulses. The values of the other parameters are: $D=0.1$, $A=0.1$, $N=10^3$, $P=10^6$, $\alpha=0.025$, and $T=1000$.}
\label{Fig08}
\end{figure}

Finally, we investigate the effects that a change in the pulse shape produces in the detection performances, as the waveform that results from the interaction between the photon and detector is not precisely known. 
Specifically, Fig.~\ref{Fig08}(a) shows how the $\dKC$ vs $\gamma_b$ is modified enlarging the duration of the current pulses, i.e., increasing the width $\delta t$.
We chose $A=0.1$, that is when the responses with and without the signal are almost indistinguishable. This figure well demonstrates that the performances of the detection improve considerably as $\delta t$ increases. 
We have also investigated the effects due to a rounding of the current pulses forming the signal feeding the junction. 
To model this type of excitation, we consider in Eq.~\eqref{gamma_ac} a combinations of arctangent instead of the theta functions. In particular, we have chosen a pulse centered in $t_0$ and having a width $\delta t$, written as $\arctan \left [ \epsilon \left ( t-t_0+\delta t/2 \right ) \right ]-\arctan \left [ \epsilon \left ( t-t_0-\delta t/2 \right ) \right ]$, where the parameter $\epsilon$ gives the rounding of the pulse corners (so that the smaller $\epsilon$ the smoother the steps). 
Figure~\ref{Fig08}(b) shows the results obtained by considering pulses of width $\delta t=200$ for the parameter $\epsilon=\left\{0.5,2,10\right\}$.
The results confirm that the proposed detection method is quite robust even in the presence of a train of rounded pulses, for the effect of the smoothing is quite marginal, being a rounded pulse a little more difficult to detect only in the $\gamma_b$ region of values where the $\dKC$ is highest.

%


\section{Conclusions}
\label{Sec04}\vskip-0.2cm
In summary, we have shown that the distributions of escape times can be effectively employed to determine the presence of pulsed signals affecting a metastable Josephson junctions system. 
In particular, we have investigated the response of an underdamped current-biased Josephson junction, embedded in a thermal noise background, to a train of current pulses that mimics a weak microwave photon field. 
We compare the distributions of mean switching times in the case in which the photon signal is absent, so that only thermal-induced processes are recorded, and where the combined action of thermal noise and current pulses can speed-up the escape process. 
To efficiently analyze these conditions in the framework of signal detection, we exploited a statistic tool that estimates the SNR, the Kumar-Caroll index, through which one can estimate how the performance of the detector depends on the system parameters.

We also define two alternative KC indexes, the ordinary $\dKC$ and the binomial $\tdKC$, that refer to two different methods, one in which the average switching times in the two settings (with and without signal) are compared and the other in which the number of switching events in the two conditions are compared. 
In general, for a given set of parameters one of the two $\dKC$ is not defined (for example, the conventional index $\dKC$ is not properly defined if not all events occur within the maximum observation time $P$), and therefore the other method should be used, whose goodness is quantified by the KC index. If there were parameter regions where both methods are operational, then the higher $\dKC$, i.e. the signal-to-noise ratio, would indicate the better method.
In both cases, the indices give a prescription of the effectiveness of the method for the number of events considered.
In this way it is possible to estimate the number of data to be recorded to achieve the desired statistical accuracy of the detection, inasmuch Eq.(\ref{variance_mean}) entails that the detection improves with the square root of the number of events \cite{Pie21}.

The $\dKC$ index, for each value of bias current, temperature, and signal amplitude, highlights the conditions that give an effective detection, that is the combination of junction parameters for which the KC index is maximized.
It is important to stress that the KC index represents one possibility among others~\cite{Kumar84,Mah36,Mac05} to summarize with a single value the possibility to discriminate between two distinct hypotheses.
For instance, a similar statistical estimator, called Hellinger distance~\cite{Spe14}, has been employed to distinguish switching current distributions in a junction operating as a qubit for microwave photon detection~\cite{Oel17}.

This work is focused on the junction details, without prescriptions neither for the dynamics of the cavity absorbing the photons neither for the transmission line specifics~\cite{Oel18}.
Instead, we concentrate on the methodological aspects of the analysis of the switching time distributions for a suitable photon detection. However, limitations in the fabrication process or in the electrodynamics of the cavity can restraint the detection performances, for instance increasing the effective noise.

Finally, the detection method discussed in this work will be employed for the analysis of data collected from single microwave photons detectors under study within the SIMP project~\cite{AlesiniBabusci20,Alesini20,Gat21}. Specifically, current-biased JJs are tested at Frascati National Laboratory (LNF) as switching detector~\cite{Kuzmin18}, and, recently, a chip fabricated at the CNR-IFN in Rome with Al transmission lines terminated by a single JJ or a dc-SQUID was tested at $10\;\text{mK}$ to verify the sensitivity to rf photons.

\section*{Acknowledgments }
The authors wish to acknowledge financial support from Italian National Institute for Nuclear Physics (INFN) through the Project SIMP and from University of Salerno through projects FARB19PAGAN, FARB20BARON.


%

\end{document}